\documentclass[12pt]{article}
\usepackage{latexsym, epsfig}

\renewcommand{\baselinestretch}{1.0}
\newcommand{\A}{\mbox{$\bf A$}}
\newcommand{\q}{\mbox{$\bf q$}}

\newcommand{\I}{\mbox{${\bf I}_{J \times J}$}}

\newcommand{\uu}{\mbox{$\bf u$}}

\newcommand{\xj}{\mbox{$\xi_j^2$}}
\newcommand{\xk}{\mbox{$\xi_k^2$}}
\newcommand{\nj}{\mbox{$\nu_j$}}

\newcommand{\hk}{\mbox{$|\bar{h}_k|^2$}}
\newcommand{\hkl}{\mbox{$|\bar{h}_{kl}|^2$}}
\newcommand{\hjl}{\mbox{$|\bar{h}_{jl}|^2$}}
\newcommand{\hj}{\mbox{$|\bar{h}_j|^2$}}

\newtheorem{theorem}{Theorem}

\begin{document}

\renewcommand{\baselinestretch}{1.0}
\title{{\Large \bf Capacity Regions and Optimal Power Allocation for Groupwise 
Multiuser Detection \footnote{This research was supported in part by the 
National Science Foundation under Grant CCR-99-79441 and by 
the New Jersey Center for Pervasive Information Technology. This work was presented in part at the 2003 Conference on Information Sciences and Systems, The Johns Hopkins University, Baltimore, MD, March 2003.}}}
\author{Cristina Comaniciu\footnote{Stevens Institute of Technology, ccomanic@stevens-tech.edu. This work was completed while the author was with Princeton University.} \,\,and H. Vincent Poor\footnote{Princeton University, poor@princeton.edu}}
\date{}

\maketitle
\thispagestyle{empty}

\begin{abstract}

In this paper, optimal power allocation and capacity regions are derived for
GSIC (groupwise successive interference cancellation) systems operating in  
 multipath fading channels, under imperfect channel estimation conditions.
It is shown that the impact of channel estimation errors on the system 
capacity is two-fold: it affects the receivers' performance within a group of 
users, as well as the cancellation performance (through cancellation errors).
An iterative power allocation algorithm is derived, based on which it can be
 shown that that the total required received power is minimized when the 
groups are ordered according to their cancellation errors, and the 
first detected group has the smallest cancellation error.

Performace/complexity tradeoff issues are also discussed by directly 
comparing the system capacity for different implementations: GSIC with 
linear minimum-mean-square error (LMMSE) receivers within the detection 
groups, GSIC with matched filter receivers, multicode LMMSE systems, and 
simple all matched filter receivers systems.  

\end{abstract}

\pagebreak

\setcounter{page}{1}
\renewcommand{\baselinestretch}{1.0}
\vspace{-.3cm}
\section {Introduction} 

Groupwise multiuser detection \cite{prasad_gr} has recently emerged as an 
appealing solution for multirate multiuser detection, since it allows for 
interference cancellation in groups, and the groups can be straightforwardly 
formed by considering users that have equal transmission rates. A natural 
detection order has been proposed in the literature \cite{wit_oj}, which 
considers the detection of the high rate users first. These high rate users 
are expected to cause more interference due to high power requirements, and 
in turn, to be less sensitive to the low power users' interference.
Within a group, any type of detectors can be implemented, although the 
simplest, most common choice is to use matched filter receivers.
Groupwise successive interference cancellation (GSIC) performance analyses 
and iterative power control schemes have been presented in \cite{bambos_gr} 
for a simplified case that considers perfect interference cancellation 
among groups and matched filter receivers within groups. However, the effect 
of interference cancellation errors has been only illustrated in \cite{bambos_gr} using simulations, and no general insight 
into the system performance can be gained without analytical results.

In this paper, we analyze the performance of a power controlled large scale
GSIC system with linear minimum-mean-square error (LMMSE) detectors within a 
group, in a multipath fading environment. Optimal power allocation is 
determined under the assumption that the channel is not perfectly known, but 
an estimate of the channel gain and of the estimation error variance can be 
supplied by the channel estimator.
We consider that the impact of channel estimation errors is two-fold:
it affects the receivers' performance within a group of users as well as
the cancellation performance through cancellation errors.

The paper is organized as follows. Section 2 presents the system model.
Section 3 discusses the optimal power allocation and illustrates capacity
regions for a two class system; performance/complexity issues are also 
presented; optimal detection ordering for interference reduction is also
discussed. Finally, Section 4 presents our conclusions.

\vspace{-.3cm}
\section{System Model}

We consider a large scale  power controlled groupwise MMSE system, in which 
the users having the same transmission rates are grouped together and decoded
using LMMSE receivers. All signature sequences are independent, randomly 
chosen, and normalized, and different transmission rates are achieved for 
different classes of users by using different spreading gains $N_j$, 
$j=1, \ ..,J$.  For simplicity, each group of users has the same target
signal to interference ratio (SIR) $\gamma_j$, although the analysis can be 
extended to allow for a more general case in which multiple target SIR 
choices are available within a group.

The first detected group is selected according to an optimal
detection criterion such as minimum received power. Then,
the interference caused by the first group is reconstructed and cancelled
from the received signal. This is done successively until the last group of
users has been detected. 

It is assumed that, due to fast fading, the channel cannot be perfectly 
known, and it is characterized by its estimated average link gain  
$\bar{h}_j$ and its estimation variance ${\xj}$, both of which we assume to be the same for all users from an arbitrary class $j$. In fact, it is reasonable
to assume that they are equal for all users, since the channel model 
considered in the analysis is in fact conditioned on the slower fading 
(free space path loss and shadow fading), which does not affect the received 
power over the time scale of interest. The effects of path loss and shadow
fading can be considered separately and mitigated by implementing power 
control loops.
Consequently, for our model, the effects of slow fading can be absorbed into 
the attenuated transmitted power, defined as
\begin{equation}
P_k = z_kP_k^t , \  k=1,\ 2,\ ..., \ K,
\label{eq:att} 
\end{equation}
\noindent where $z_k$ is the path loss due to free space loss and shadow 
fading, $P_k^t$ is the transmitted power and $K$ is the total number of users 
in the system.

For a multipath fading channel with $L$ resolvable 
paths, we denote by $\hj$ the equivalent estimated average power gain 
defined as
\begin{equation}
\hj = \sum_{l=1}^{L}\hjl, \ j=1,\ 2,\ ..., \ J,
\end{equation}
\noindent where $\bar{h}_{j,l}$ represents the average link gain for the 
$l^{th}$ path for a user in class $j$.

The imperfect channel estimation yields an imperfect cancellation for 
group $j$ of users, resulting in a residual interference power 
$\sum_{i=1}^{K_j}\epsilon_j Q_{j,i}$, where $K_j$ is the number of users in 
 class $j$, $Q_{j,i}=P_{j,i}\hj$ is the received power of user $i$ from 
class $j$, and $\epsilon_j$ is the fractional error in canceling the total 
interference power created by the $j^{th}$ group. This implicitly assumes 
that the fractional error for canceling a group $j$ user is the same for all 
users in class $j$. 
Since the target bit-error rates (BERs) are usually very low, it can be 
assumed that the cancellation error is mostly determined by the amplitude and 
phase estimation errors. Similarly to the approach in \cite{And:opt}, we 
assume that the cancellation error ($\epsilon$) for the successive 
interference cancellation is approximately the same as the total channel 
estimation standard deviation, $\xi$. Assuming further that the multipath 
fading components are i.i.d. (independent and identically distributed) and 
have estimation error variances of $\xi^2$, the estimated cancellation error 
for an $L$ path channel can be approximated by:
\begin{equation}
\epsilon \simeq \sqrt{L\xi^2}.
\end{equation} 
The estimation error variances can be determined as in \cite{evtse}.

\vspace{-.5cm}
\section{Optimal Power Control and System Capacity} 
\vspace{-.3cm}
\subsection{Optimal Power Allocation}

It has been shown in \cite{evtse} that the achieved SIR for a large CDMA 
system (number of users and the spreading gain both increase
without bound while their ratio is a constant $\alpha=K/N$) using LMMSE 
receivers can be expressed as
\begin{equation}
SIR_k = \frac{P_k \sum_{l=1}^L\hkl \beta}{1+P_k \xk \beta} = \frac{P_k 
\hk \beta}{1+P_k \xk \beta},
\label{eq:grSIRfad}
\end{equation}
\noindent where $\beta$ is the unique fixed point in $(0, \infty)$ that 
satisfies
\begin{equation}
\beta=\left[\sigma^2+ \frac{1}{N}\sum_{k=2}^K \left( (L-1)I(\xk P_k, \beta)+
I(P_k(\xk+\hk),\beta)\right) \right]^{-1}, 
\label{eq:grbetafad2}
\end{equation}

\noindent where $I(p, \beta)=\frac{p}{1+p\beta}$, and $\sigma^2$ is the 
background noise power.

Based on the SIR expression in (\ref{eq:grbetafad2}), it can be shown 
(see \cite{com_sp}) that, for equal channel characteristics, all users 
having the same transmission rate (in the same detection group) and same SIR 
requirement must have equal received powers. 
For our GSIC system, we denote the groups as $1, 2, \ ..., J$, which 
represents the detection order, and we assume that within a group, all users
are detected using LMMSE receivers. It can be shown that every 
group $j$ of users can be approximated as an all LMMSE system with enhanced 
noise $\Sigma_e^j$:
\begin{equation}
\Sigma_e^j = \sigma^2 + \sum_{l<j}\sum_{k=1}^{K_l}\frac{1}{N_l}\epsilon_lQ_l + 
\sum_{l>j}\sum_{k=1}^{K_i}\frac{1}{N_l}Q_l= \sigma^2 + \sum_{l<j}\epsilon_l
\alpha_lQ_l + \sum_{l>j}\alpha_lQ_l. 
\label{eq:sigmaeqgr}
\end{equation}

This equivalence is based on the fact that the receiver filter coefficients
for group $j$ users ignore the structure of the interference from other 
groups, and thus any pair of filter coefficients and signal signature 
sequences for users in other groups may be considered to be independent. 
Based on (\ref{eq:grSIRfad}) and (\ref{eq:sigmaeqgr}) and following a similar 
reasoning as in \cite{com_sp}, a power control feasibility condition
for the GSIC system can be derived as follows.

\noindent From (\ref{eq:grSIRfad}), we can express $\beta$ such that a class 
$j$ of users can meet their target SIR $\gamma_j$:
\begin{equation}
\beta \geq \frac{\gamma_j}{Q_j(1-\nu_j\gamma_j)}, \ j=1, \ ...,J,
\label{eq:betagr}
\end{equation}

\noindent with $\nj = \frac{\xj}{\hj}$.
Since $\beta$ is required to be positive, a feasible  target SIR is 
obtained if  $\gamma_j < 1/\nu_j$.

Using (\ref{eq:grbetafad2}), (\ref{eq:sigmaeqgr}) and (\ref{eq:betagr}), and 
after straightforward algebraic manipulation, we can derive the power 
feasibility condition such that target SIR $\gamma_j$ can be met with 
equality for an arbitrary class $j$ of users:
\begin{equation}
Q_j=\theta_j \sum_{l<j}\epsilon_l\alpha_lQ_l + \alpha_jQ_j\Lambda_j+ \theta_j \sum_{l>j}\alpha_lQ_l + \theta_j  \sigma^2,
\label{eq:SIRreqgr}
\end{equation}

\noindent where $\theta_j=\gamma_j/(1-\nu_j\gamma_j)>0,$ and $\Lambda_j=(L-1)\nu_j \gamma_j + (1+\nu_j)\gamma_j/(1+\gamma_j)$.

Given that target SIRs have to be met for all users, the power control 
feasibility can be expressed as a matrix equation condition
\begin{equation}
(\I-\A)\q =\sigma^2 \uu,
\label{eq:SIRmeqgr}
\end{equation}
 
\noindent where $\q^T= [Q_1, Q_2, \ ..., Q_J]$, $\uu^T= [\theta_1, \theta_2, \ ..., \theta_J]$, $\I$ is the identity matrix, and 
\begin{equation}
\A = \left(\begin{array}{c} \alpha_1\Lambda_1 \\ \epsilon_1 \alpha_1 \theta_2 \\ \mbox{ ... } \\ \epsilon_1 \alpha_1 \theta_J  \end{array} \begin{array}{c}  \theta_1 \alpha_2  \\ \alpha_2\Lambda_2 \\ \mbox{ ... } \\ \epsilon_2 \alpha_2 \theta_J \end{array} \begin{array}{c}   \mbox{ ... }  \\  \mbox{ ... } \\ \mbox{ ... } \\  \mbox{ ... } \end{array} \begin{array}{c} \theta_1 \alpha_J  \\ \theta_2 \alpha_J \\ \mbox{ ... } \\  \alpha_J \Lambda_J \end{array}\right). 
\label{eq:Amatrix}
\end{equation}

The matrix $\A$ is a nonnegative matrix, but it is not necessarily 
irreducible, since a perfect cancellation for group 1 users results in a 
reducible matrix. 
For a nonnegative, irreducible matrix, a positive vector solution to 
(\ref{eq:SIRmeqgr}) exists iff $\rho(\A)<1$, where $\rho(\A)$ is the spectral
radius of $\A$. This is usually the practical case since perfect cancellation
is hard to achieve. Nevertheless, using similar arguments as in \cite{andrews}, it can be shown that the above result still holds for matrix  $\A$ even
though it is not irreducible. The power control feasibility result can be
summarized in the following theorem.

{\theorem In a groupwise successive interference cancellation system with 
LMMSE receivers within a group, and operating under a multipath fading 
environment with imperfect channel estimation, a positive power vector 
solution exists such that all users meet their target SIRs $\gamma_j$, if 
and only if
\begin{equation}
\gamma_j < \frac{1}{\nu_j} \mbox{ and } \rho(\A)< 1;
R\label{eq:condgr}
\end{equation}

\noindent The optimal received power allocation for the groups of users is 
given by
\begin{equation}
\q^*=(\I-\A)^{-1}\uu \sigma^2.
\end{equation}
}
Distributed, iterative power control algorithms based on the GSIC system 
can be implemented as
\begin{equation}
\q^*(n)=i(\q^*(n-1)),
\label{eq:stdpcf}
\end{equation}
\noindent where $n$ is the current iteration number, and $i(\q^*(n-1))$ is a 
standard interference function, computed as a function of the powers 
at iteration $n-1$. Since (\ref{eq:stdpcf}) is expressed using a standard 
interference function, it can be proven that it converges to a minimum power 
solution for both synchronous and asynchronous updates \cite{yates_fwk} if 
the power control is feasible.   

\vspace{-.3cm}
\subsection{Optimal Detection Order}

It can be shown that the received power requirements 
for different groups can be derived using a recursive formula 
\cite{com_ciss03}. 
Denoting by $Q_i$ the required received power for detection class $i$, and using the notation $\Gamma_i=(1-\alpha_i\Lambda_i)/\theta_i$,
\begin{equation}
Q_{i+1}=\frac{\Gamma_i+\epsilon_i\alpha_i}{\Gamma_{i+1}+\alpha_{i+1}}Q_i,
\label{eq:rel}
\end{equation}

\noindent or equivalently,
\begin{equation}
Q_{i}=\displaystyle{\Pi_{j=1}^{i-1}\frac{\Gamma_j+\epsilon_j\alpha_j}{\Gamma_{j+1}+\alpha_{j+1}}Q_1},
\label{eq:rel2}
\end{equation}
\noindent with $\displaystyle{Q_1=\frac{\sigma^2}{\Gamma_1-\sum_{i=2}^J\alpha_i\displaystyle{\Pi_{j=1}^{i-1}\frac{\Gamma_j+\epsilon_j\alpha_j}{\Gamma_{j+1}+\alpha_{j+1}}}}}$.

\noindent The total received power requirements for all users, for a given 
detection order, can then be express as
\begin{equation}
Q_T= \sum_{i=1}^J \alpha_iQ_i = Q_1(\alpha_1+\Gamma_1)-\sigma^2;
\label{eq:rec}
\end{equation}
\noindent that is,
\begin{equation}
Q_T= \frac{\sigma^2}{\frac{\Gamma_1}{\alpha_1+\Gamma_1}-\frac{1}{\alpha_1+\Gamma_1}\sum_{i=2}^J \alpha_i \displaystyle{\Pi_{j=2}^{i}\frac{\Gamma_{j-1}+\epsilon_{j-1}\alpha_{j-1}}{\Gamma_{j}+\alpha_{j}}}}-\sigma^2.
\label{eq:rec2}
\end{equation}

While the above results were derived for a given, arbitrary,
detection order, this can be optimized for a minimum received power 
solution. Using a similar approach to that in \cite{niu}, it can be shown 
\cite{com_ciss03} that $Q_T$ is minimized if the groups are ordered with 
respect to their cancellation errors, with the first group detected being 
the one with the lowest cancellation error. 

An interesting observation is that this result may be in contrast with the 
popular recommendation of detecting higher rate users first. Although the 
present analysis  considers only the impact of the imperfect 
amplitude estimation on the cancellation errors, this model can be extended to also encompass other effects, such as the performance differences 
between the asymptotic analysis and the practical finite case. In this case, 
higher rate users (using lower spreading gains)
may have a higher cancellation error due to a higher achieved SIR 
variance relative to the estimated average SIR for asymptotically large 
systems \cite{evtse}. 

\subsection{Capacity Considerations}

Capacity regions for the general case of a GSIC system with $J$ groups
can be defined in a generic form as
\begin{equation}
\mbox{{\large \it C}} = \left\{(\alpha_1, \alpha_2, ...\  , \alpha_J) \ |\  
\gamma_j<1/\nu_j, \ \forall j=1, \ ..., J, \  \rho(\A)<1 \right\}.
\end{equation}

The computation of the maximum eigenvalue $\rho(\A)$ is not 
very complex since $\A$ is a $J \times J$ matrix, where $J$ is the 
number of groups, which is usually a small number due to detection delay 
constraints.

For the particular case of a GSIC system with two detection groups, an
explicit dependence between the number of users that can be supported
in each class can be obtained as
\begin{equation}
\alpha_1\Lambda_1 + \alpha_2 \Lambda_2 + \sqrt{\Delta} < 2,
\label{eq:capacgr}
\end{equation}

\noindent where $\Delta = (\alpha_1\Lambda_1+\alpha_2\Lambda_2)^2+4\alpha_1\alpha_2(\theta_1 \theta_2 \epsilon_1 - \Lambda_1\Lambda_2)$.

The capacity in (\ref{eq:capacgr}) represents a performance benchmark, as it
combines the advantages of GSIC and linear MMSE receivers. A more practical
case (less complex implementation) would be the GSIC with matched filter 
receivers within groups, which would still give performance improvements 
compared with the case with no interference cancellation. The capacity for 
GSIC with matched filter receivers can be derived similarly to the previous
derivation for GSIC with LMMSE, starting from (\ref{eq:grSIRfad}) and 
(\ref{eq:grbetafad2}), with $I(p)=p$. For the 
particular case of a two class system, the capacity formula is given by
(\ref{eq:capacgr}), with $\Delta$ replaced by $\Delta^* = (\alpha_1\Lambda_1^*+\alpha_2\Lambda_2^*)^2+4\alpha_1\alpha_2(\theta_1 \theta_2 \epsilon_1 - \Lambda_1^*\Lambda_2^*)$, and $\Lambda_j$, $j=1,2$, replaced by $\Lambda_j^*= \frac{\gamma_j}{1- \nu_j\gamma_j}(L\nu_j+1)$, $j=1,2$.

\begin{figure}
\centerline{
\epsfxsize=3.5 in\epsffile{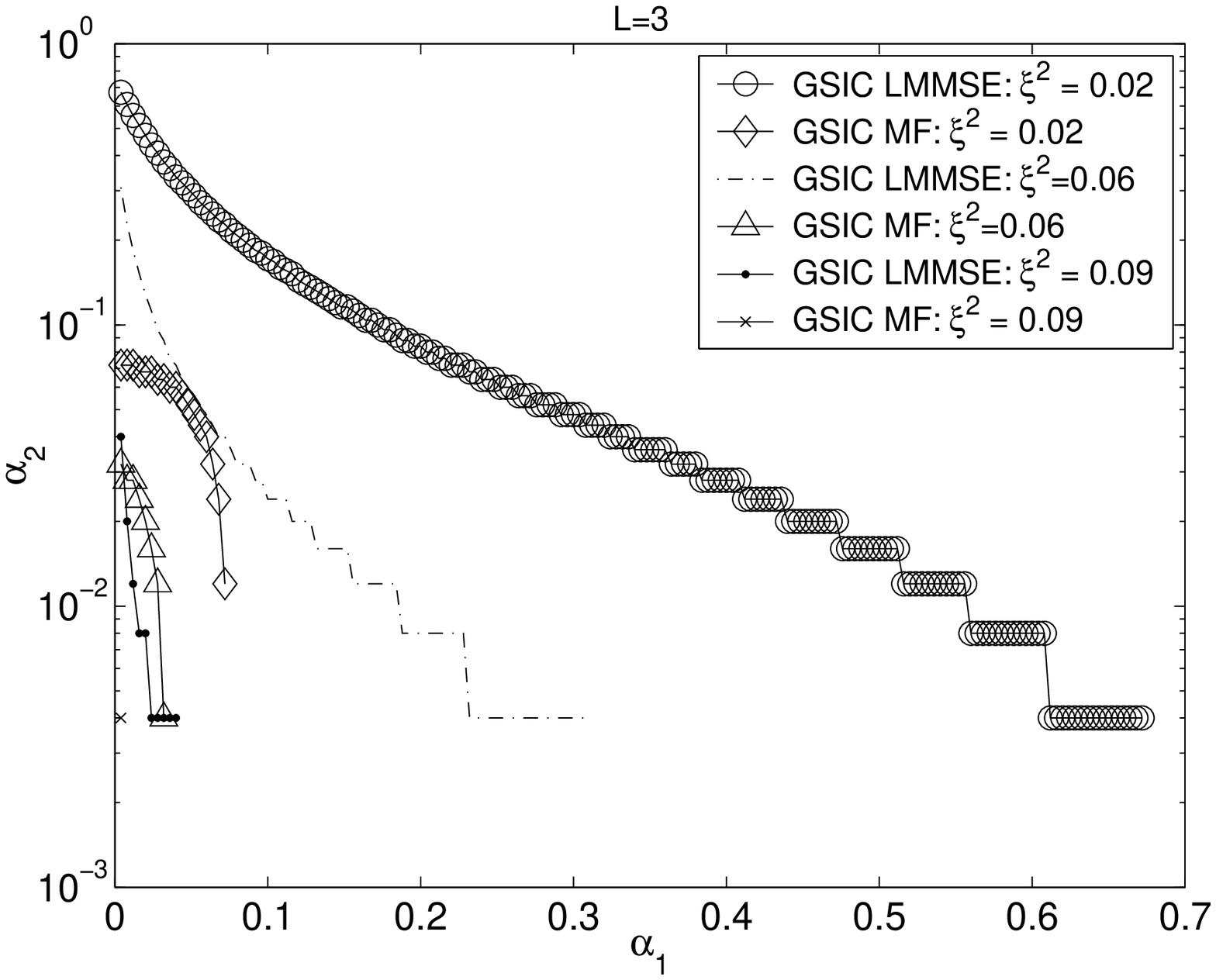}\hspace{1.0pc}}

\caption{Capacity comparisons: GSIC with LMMSE versus GSIC with MF}
\label{fig:gsic2}
\end{figure}

\begin{figure}
\centerline{
\epsfxsize=3.5 in\epsffile{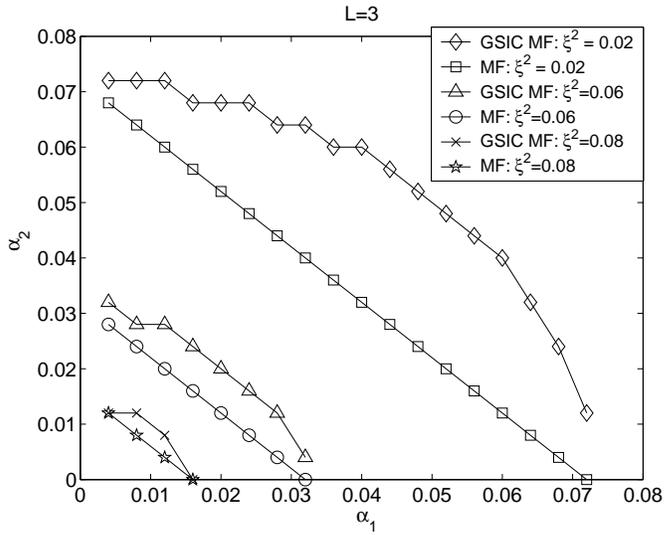}\hspace{1.0pc}}

\caption{Capacity comparisons: GSIC with MF versus all MF system}
\label{fig:gsic_mf}
\end{figure}

In Fig. \ref{fig:gsic2} we compare the performance of the matched filter
GSIC and the LMMSE GSIC, for different channel estimation errors and for 
required target SIRs for both classes equal to 10. The estimated average 
link gain $\hj$, $j=1,2$ is set to 1. We notice that both 
implementations are strongly affected by channel estimation errors, but a 
very substantial performance gap exists in favor of the LMMSE implementation.
This may justify the increase in implementation complexity for the GSIC LMMSE
systems for specific applications requiring high performance.
By comparing a simple all matched filter system with a GSIC system using 
matched filters within the detection groups, the advantage of the GSIC
implementation can be illustrated (see Figure \ref{fig:gsic_mf}).

\begin{figure}
\centerline{
\epsfxsize=3.6 in\epsffile{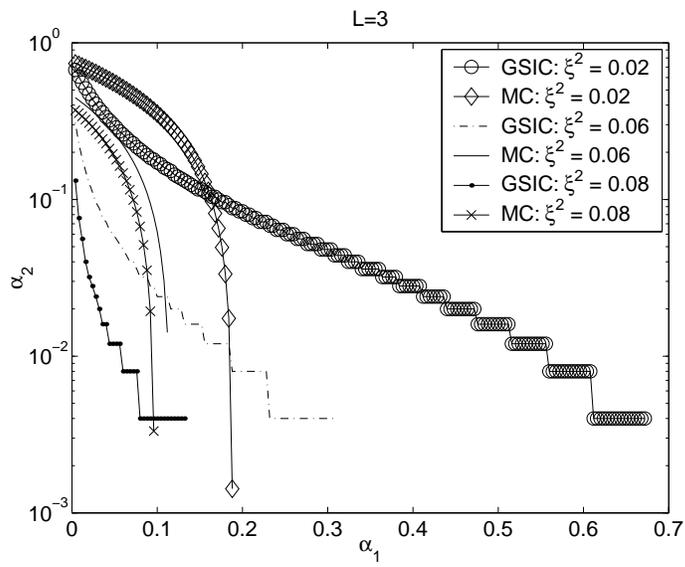}}
\caption{Asymptotic capacity regions comparison: GSIC with LMMSE versus MC LMMSE}
\label{fig:gsic1}
\end{figure}

Further, in Fig. \ref{fig:gsic1}, we also present comparisons between the 
GSIC LMMSE and an alternate multi-rate implementation: multicode LMMSE. The 
plots are obtained for channel length $L=3$, target SIRs = 10 for both 
classes, average estimated link gains $\hj = 1$, and various channel 
estimation error variances. For the multicode system, a two-class system is 
considered: class 1 is the high rate class, and class 2 the low rate one, 
with $R_1=MR_2=MR$.
Consequently, the equivalent number of users per dimension that can be
supported by the multicode system \cite{com_sp} is $(M\alpha_1,\alpha_2)$, 
and all users can meet their SIR requirements if 
\begin{equation}
M \alpha_1\Lambda_1 + \alpha_2 \Lambda_2<1,
\label{eq:capmc}
\end{equation}
\noindent with $\Lambda_i=(L-1)\nu_i \gamma_i + (1+\nu_i)\gamma_i/(1+ \gamma_i)$, $i=1,2$. 

For the numerical results, we chose $M=4$. We notice that both schemes show 
worse performance as the channel uncertainty increases. 
The multicode system performs better for a practical range for the low rate
users $\alpha_1 \geq 0.1$. However, as is well known, the multicode 
implementation has the drawback of requiring linear amplifiers. 

\section {Conclusions}

In this paper, we have studied optimal power allocation and the capacity
regions of an LMMSE GSIC system in multipath fading channels for 
asymptotically large systems. We have assumed that,
although the channel cannot be perfectly estimated, channel estimation 
error statistics are available, and are quantified by the estimation error 
variance.

We have shown that the impact of channel estimation 
errors is two-fold: it impacts the LMMSE receiver performance within a 
class of users in the same detection group, and also it is strongly related 
to the cancellation errors for the successive group interference 
cancellation.  
We have also shown that, similarly to successive interference 
cancellation (SIC) systems \cite{andrews, niu}, the optimal 
ordering implies that the groups should be detected according to their 
cancellation accuracy, that is, the ones with the highest error rates 
should be detected last. This yields a minimum total required received power, 
which in turn reduces the required transmission power resulting in reduced 
inter-cell interference seen by the neighboring cells.

Performance/complexity tradeoffs for various implementation scenarios 
involving GSIC, LMMSE and matched filter receivers have also been presented.

\vspace*{-.5cm}
\bibliography{C:/Cristina/GRWMMSE/book1}
\bibliographystyle{abbrv}

\pagebreak

\end{document}